\documentstyle[a4,12pt,epsfig,graphicx]{article}
\begin{document}
\titlepage
\begin{flushright}
CERN-TH/99-391\\
t99/140\\
hep-th/9912103 \\
\end{flushright}
\vskip 1cm
\begin{center}
{ \Large
\bf The Supermoduli Space of Matrix String Theory}
\end{center}

\vskip 1cm
\begin{center}
{\large Ph. Brax\footnote{email: philippe.brax@cern.ch} }
\end{center}
\vskip 0.5cm
\begin{center}
Theoretical Physics Division, CERN\\
CH-1211 Geneva 23\footnote { On leave of absence from  Service de Physique Th\'eorique, 
CEA-Saclay F-91191 Gif/Yvette Cedex, France}\\
\end{center}

\vskip 2cm
\begin{center}
{\large \bf Abstract}
\end{center}
\vskip .3in \baselineskip10pt{
We study   matrix string scattering amplitudes and matrix string instantons  on a marked  Riemann
surface in the limit of a
vanishing string coupling constant.
We give an explicit parameterization of
the moduli space of such instantons.  We also give a 
description of the set of fermionic supermoduli. The integration over the supermoduli leads to the
inclusion of  picture changing operators at the interaction points.
Finally we investigate the large $N$ limit of the measure on the instanton
moduli space and show its convergence to the Weil-Petersson measure
on the moduli space of marked Riemann surfaces.  
}
\bigskip
\vskip 1 cm

\noindent
\newpage
\baselineskip=1.5\baselineskip
\section{Introduction}
Matrix string theory\cite{dvv}\footnote{DVV in the following.} is a
 concrete proposal for a non-perturbative definition
of type IIA superstring theory\cite{dhokerphong}. 
The basic building blocks of this model are bosonic and fermionic matrices in the adjoint representation of $U(N)$.  The
 link with string theory in the light cone gauge is provided by
 the analysis  of the moduli space of vacua showing that one retrieves the light cone 
Green-Schwarz action in the small string coupling limit. Due to the matrix nature of the model,
one gets $N$ copies of the Green-Schwarz action.

It has soon been realized that string interactions\cite{dvv,tom} can also be incorporated in the matrix context. 
 In particular one finds that the 
string interactions are represented by Mandelstam diagrams\cite{mandel} where the propagation of strings along
 straight lines ends up at an interaction point where two strings join up. 
This picture emerges in the study of the spectral cover
of the cylinder, i.e. 
the $N$ sheets formed by the eigenvalues of the scalar field $X=X_1+iX_2$.
The  description of the role of the spectral cover in  the string interactions has been first given in \cite {tom} and refined in \cite{nesti1,nesti2}. In the latter it has been shown that the light-cone fields describing the Green-Schwarz action in the small string coupling limit live on the world sheet formed by the spectral cover. The upshot being that in the small string coupling limit the matrix string theory reduces to the light-cone Green-Schwarz theory on the spectral cover supplemented with a residual $U(1)$ theory.

These properties have been deduced from the analysis of the matrix string instantons\cite{vergi}. 
The Yang-Mills action describing matrix string theory admits instantonic solutions which correspond
 to the Euclidean version of BPS configurations.
These instantons
 are solutions of a Hitchin system of differential equations\cite{hitch1,hitch2,hitch3}.
The Mandelstam diagram representing the string interaction corresponds to the spectral cover of the Hitchin system. This provides a direct link between the Yang-Mills and string points of view. 

One would like to obtain an explicit description of the matrix string instanton moduli space ${\cal M}$.  An argument
 based on an index theorem\cite{nesti3} shows that the moduli space
has complex dimension $(2g_S-3+p)$ where $g_S$ is the genus of the spectral
cover $S$  and $p$ is the number of external states involved in a scattering amplitude. Moreover the authors
 of \cite {nesti3} argue that there are  $2g_S$ further discrete coordinates specifying the scattering
 amplitudes in  matrix string theory. This  identifies the moduli space of instantons with a discrete slice of
 the moduli space of marked Riemann  surfaces .

 A complete identification of the scattering amplitudes in type IIA superstring theory and matrix string
theory requires the equality between the measures on the moduli spaces. Due to the discretization in matrix
 theory this can only be valid in the large $N$ limit.

Recently it has been argued that matrix string on the torus has a discrete moduli space whose large $N$ limit
 is exactly the toroidal moduli space of string theory\cite{kostov}. This strongly suggests that the large $N$ limit
 has to be taken in  order
to retrieve the string measure on the moduli space of Riemann surfaces. 

In superstring theory in the light cone frame one includes an operator at the interaction points whose role is to guarantee the ten dimensional Lorentz invariance. An  argument given in DVV suggests that the interaction points of strings in matrix theory should be decorated
 with the insertion of an operator playing the same role as the picture changing operators in the RNS
 version of string theory\cite{verlin2}.
It is known in string theory that these operators appear as the result of the integration over
 the super-moduli\cite{verlin3}. One of the aims of this paper is to deal with this issue.

In this work we study the explicit solutions to the instanton equation on an arbitrary marked 
Riemann surface. This is equivalent to finding a  family of flat bundles over a marked Riemann surface
 together with a reality restriction on the solutions.  
In section II
we define a $(2g_S-3+p)$ family of solutions, the moduli space of instantons.
 Geometrically these instantons are in one to one correspondence with the family  of spectral
 covers characterized by the zero set of a meromorphic one form with poles at the marked points.
 This provides an embedding of the instanton moduli space into the moduli space of marked
 Riemann surfaces. In section III we give a stringy description of this moduli space
 of instantons in terms of the string scattering diagram associated to the spectral cover.
  The zeros of the meromorphic one-form  on the spectral cover are   identified with the
 interaction points\cite{dhoker2,gidd} of the string diagram. This construction provides the backbone to the
 dynamics of the string interactions. The kinematics, i.e. the momenta, are given
 by the discrete Wilson lines that one can allow along non-trivial cycles of the spectral cover when constructing flat bundles.
This completes the identification of the moduli space of instantons as a discrete slice
  of the moduli space of marked Riemann surfaces.
In the formal large $N$ limit the two moduli spaces coincide. 
In section IV we review the calculation of the scattering amplitude and the lifting of fields to the spectral cover. 
 In section V we  study the issue of the fermionic super-moduli and show that they are intimately linked with
 fermionic ghost zero modes associated to the interaction points of the string diagram. 
We perform the 
 integration over the supermoduli and retrieve the DVV prescription  where a picture changing operator
is included at each interaction point.  We also mention the ambiguities resulting in the integration over the super-moduli.
Finally we  consider the large $N$ limit of the scattering amplitude showing that the measure on the moduli space of
 instantons converges weakly to the Weil-Petersson measure on the moduli
space of marked Riemann surfaces. This shows that the scattering amplitudes in matrix string theory and string theory
coincide in the small coupling limit.

\section{Instantons on  Marked  Riemann Surfaces}
In this section we  study the small string coupling 
regime of matrix string theory. Matrix string theory is obtained
by dimensional reduction of the ten dimensional $U(N)$ super Yang-Mills
theory on a cylinder. 
The ten dimensional super Yang-Mills theory is endowed with a gauge field $A$ and a 
space-time fermion $\Theta$. Upon reduction to two dimensions the gauge field splits into a two dimensional gauge field and
eight bosonic coordinates.
The two
dimensional action of matrix string theory is given by the dimensional reduction
\begin{equation}
S=-\frac{1}{2\pi}\int d\tau d\sigma \hbox {Tr} [  (D_{\alpha}X^{I})^2 
-i\Theta^T\gamma^{\alpha}D_{\alpha}\Theta 
+{1\over 2g^2}F^2_{\alpha\beta}+{g^2\over 2}[X^I,X^J]^2
-i\Theta^T\gamma_I[X^I,\Theta]].
\label {SYM}
\end{equation}
The fields are $N\times N$ Hermitean matrices. The index $I$ runs 
from 1 to 8 and the sixteen fermions split into the $8_s$ and $8_c$ 
spinorial representations of $SO(8)$. The string coupling constant 
of the type IIA string theory is $g_s$ such that $\alpha 'g_s^2=g^{-2}$ where $g$ is the Yang-Mills coupling constant. The coordinate $\sigma$ 
lives between 0 and 2$\pi$.
All the fields are world sheet scalars subject to a periodicity
condition in the $\sigma$ direction. The fermions are periodic due to the  original space-time supersymmetry. 
In the following we will mostly be
interested in the Euclidean version of this action obtained after
a Wick rotation on the cylinder.

The path integral defining correlation functions in the matrix model
is  dominated by instanton configurations in the small 
string coupling limit $g_s\to 0$. A semi-classical evaluation of
the path integral is then available leading to a link with
the string scattering amplitudes\cite {dvv,tom,nesti2}. The classical instantons
are Euclidean versions of BPS solitons preserving half of the supersymmetries which are
 derived  from the dimensional reduction of the SYM ten dimensional
supersymmetry transformations. 
Imposing that the vanishing of the spinors  $\Theta$ is consistent with 
 half of the supersymmetry requires that  the gauge field and the scalars satisfy 
 the Hitchin system \cite {hitch1,hitch2,hitch3}
\begin{eqnarray}
F_{w\bar w}+ig^2[X,\bar X]&=&0\nonumber\\
D_{\bar w}X&=&0\nonumber\\
\label{in}
\end{eqnarray}
where $w=\frac {1}{2}(\tau-i\sigma),\ X= X^1+iX^2$ and $A_w=A_0+iA_1$. The
covariant derivative acts in  the adjoint representation of the
gauge group $U(N)$.
The solutions to these equations have been extensively studied\cite{nesti1,vergi}. In particular it is known that 
the gauge configuration is almost flat $F_{w\bar w}\approx 0$ away from the core of the instantons- 
a neighbourhood of a finite set of points-
where the intrinsically non-commutative nature of the instantons                             
is  apparent. Moreover it is in the core that supersymmetry is broken by the instantons.
In the large YM coupling limit $g\to \infty$ the size of the core of the instantons vanishes leaving
an almost everywhere flat gauge configuration.

In the following we shall be interested in recovering the type IIA string perturbation
results from matrix string theory. To do so we will concentrate on the flat part of the 
instantons  away from their core. In the very small string coupling limit these field configurations
are defined 
 on the whole cylinder where a few points have been singled out. These points will
 happen to be branched points where the gauge configuration is ill-defined. It 
is only by going to an appropriate cover -the spectral cover- that
 the gauge configuration becomes well-defined. Of course by only considering the flat
 configurations we have neglected most of the non-Abelian nature of matrix string
instantons  which is
 essential to go beyond type IIA string perturbation theory.

We shall  consider the simplified flatness  equations
\begin{equation}
F_{w\bar w}=0,\ \ D_{\bar w}X=0
\label{bundle}
\end{equation}
supplemented with the constraint
\begin{equation}
[X,\bar X]=0.
\label{real}
\end{equation}
The solutions to these equations correspond to the very small string coupling limit
of the solutions of (\ref{in}). They will be shown to be  sufficient to describe the
 nature of string interactions. Another
remarkable property of 
the  instantons (\ref{bundle}) is that they are no longer   BPS configurations.
They preserve  all the supersymmetries of matrix string theory. In particular the semi-classical expansion 
around these instantons will not lead to fermionic zero modes due to the breaking of some of the
 supersymmetries.

It is convenient to define    the form $X$ characterized locally by
the  differential $X=X(w)dw$.  The second differential equation
in (\ref  {bundle}) for $X$  is  then
cast into the form
\begin{equation}
\ \bar D_A X=0
\label {dif}
\end{equation}
where $\bar D_A$ is the covariant derivative acting on forms.
The   equations (\ref  {bundle}) can be analysed  thanks to their
conformal invariance.
This allows to map the cylinder to the
sphere with two marked points. The conformal invariance of the
equations (\ref{bundle}) allows to extend their validity to an
arbitrary Riemann surface $\Sigma$ of genus $h$ with $v$ marked points. 

 The connection
$A=A_{z}dz+A_{\bar z}d\bar z$ with values in the Lie
algebra of the gauge group $U(N)$ is flat on the Riemann surface $\Sigma$ with coordinates $z$. 
The field
$X(z)$ defines   a section of the complex  vector bundle $Adj 
P$ where $P$ is the principal vector bundle on $\Sigma$ with
fibres $Gl(N)$. Finally the real condition $[X,\bar X]=0$ imposes strong restrictions on $X$ and $A$.

Let us summarize the results obtained in the rest of this section.
The
construction of a flat $U(N)$ vector bundle with a 
section $X$ of $Adj P$ will be  carried out in two stages. 
 Notice the decomposition $U(N)=(SU(N)\times
U(1))/Z_N$ where $Z_N$ is the centre of the gauge group
acting diagonally on the two factors $SU(N)$ and $U(1)$. First we shall
construct the 
$SU(N)\times U(1)$ instantons. The explicit
solutions $(A,X)$ will be  parameterized by a moduli space ${\cal M}$
of finite dimension. Once this continuous problem has been  solved  the 
effect of modding out by the centre $Z_N$ will be  taken into account.
As usual in orbifold constructions this will necessitate to
introduce different twisted
sectors. They spring from the fact that the gauge group is not
simply connected $\pi_1(U(N))=Z_N$.

The equations (\ref{bundle}) have a complex gauge symmetry where the gauge group $G=U(N)$ is
 complexified, i.e. the form $A=A_zdz + A_{\bar z}d\bar z$ takes values in the Lie algebra of $G^c=Gl(N,C)$.
 It is easy to verify that the differential equations (\ref{bundle}) are invariant under the complex gauge transformations
 with $g\in G^c$
\begin{equation}
 A_{\bar z}\to gA_{\bar z}g^{-1} -i(\bar \partial g)g^{-1}
\end{equation}
where $X\to adj(g) X$. 
This allows to define a  decomposition of the set of solutions of
(\ref{bundle}) into complex gauge orbits. We shall show that the
bundle $Adj P$ always admits a holomorphic structure, a
holomorphic connection and a holomorphic section $X_0$ whose
complex gauge orbit possesses an instantonic configuration,
i.e. a flat unitary
connection $A$ with a section satisfying $\bar D_A X=0, \ [X,\bar
X]=0$. The resulting flat unitary connection $A$ is single-valued on the
Riemann surface $\Sigma$.  

Let us first consider the case with no marked points and give an
explicit description of the instantons.
The complex vector bundle $Adj P$ on the Riemann
surface $\Sigma$ can be endowed with a covariant derivative
$D=D'+D''$ where $D'$ sends $(p,q)$-forms to $(p+1,q)$-forms
(respectively $D''$ sends $(p,q)$ forms to $(p,q+1)$ forms). As
$(D'')^2=0$ - there are no $(2,0)$ forms on a Riemann surface- one
can always find a complex structure\cite{kob} such that
$D''=\bar \partial $. Let us now consider a holomorphic one form
$X_0$ which is also a  section
of $Adj P$ for this choice of complex structure, i.e.
$D''X_0\equiv \bar \partial X_0=0$.
The
holomorphic sections $X_0\in H^0(\Sigma, K_{\Sigma}\otimes Adj P)$, where
$K_{\Sigma}$ is the canonical bundle,  are conveniently obtained by using  
 the  Schottky representation \cite{denis}
of the   Riemann surface $\Sigma$. 
Consider a $2h$ dimensional
basis for the  homology cycles $a_i,b_i,\ i=1\dots h$ of $\Sigma$. Choose a
small cylinder $C_i$ bording the cycles
$a_i$ on each of the $h$ handles. Opening up the handles by
removing the cylinder $C_i$ creates $2h$ open discs on the
surface $\Sigma$. The open surface $\tilde\Sigma$ is the Riemann sphere
with $2h$ open discs and $v$ marked points. The discs are such that the circles on their
boundaries are associated by pairs
$(a_i^-,a_i^+)$ corresponding to the two boundaries of the
removed cylinders $C_i$.  Denote by $\gamma_i$ the homography
sending the circle $a_i^+$ onto $a_i^-$. The form $X_0$ on the circles $a^+_i$ and $a^-_i$ is related by this homography
and by the connecting matrices $H_i$ 
\begin{equation}
X_0(\gamma_i(x))\vert_{a^-_i}=adj (H_i) X_0(x)\vert_{a^+_i}.
\end{equation}
Due to these boundary conditions on the open discs the
holomorphic differential $X_0$ is given in terms of Poincar\'e series.

Let us define the following Poincar\'e series\cite{denis, nekra}
\begin{equation}
\omega_k(x_0)dz=\sum_{\gamma\in \Gamma}adj(H^{-1}_{\gamma})d\ln\frac{\gamma(z)-\gamma_k(x_0)}{\gamma(z)-x_0}
\end{equation}
where $\Gamma $ is the Schottky group. i.e. the formal product of the 
$h$ homology cycles $b_i$. To each element $\gamma$ one associates $H_{\gamma}$
as the product of the corresponding matrices $H_i$. 
The point $x_0$ is an arbitrary point on the Riemann sphere.
The holomorphic differentials  are of the form
\begin{equation}
X_0=\sum_{k=1}^h \omega_k(x_0) M_k dz
\label{poin}
\end{equation}
where $M_k$ is a matrix with values in the Lie algebra of the complexified
gauge group. Holomorphy is guaranteed once the residue of $X_0$ at $x_0$ 
vanishes, this requires
\begin{equation}
\sum_{k=1}^h (adj(H_k^{-1})-1)M_k=0
\label {mod}
\end{equation} 
The construction of the holomorphic section $X_0$ allows  to
define the instantons.

The matrix valued one form $X_0$ can be diagonalized
\begin{equation}
X_0=hX_d h^{-1}
\end{equation}
where $X_d$ is diagonal and the matrices $h$ satisfy 
 \begin{equation}
h(\gamma_i(x))\vert_{a_i^-}=H_i h(x)\vert_{a^+_i}
\end{equation}
where the $H_i$ are complex invertible matrices associated to the $b$ cycles joining two identified circles.
Notice that the matrix $h$ is holomorphic and allows to define a
holomorphic connection one form $\alpha=\alpha'+\alpha''$ 
\begin{equation}
\alpha'= -i (\partial h)h^{-1},\ \alpha''=0
\end{equation}
on the bundle $Adj P$.
The complex matrix $h$ can be factorized
\begin{equation}
h=\hat h U
\end{equation}
where $\hat h\in G^c/G$ and $U\in G$. Let us now perform the
complex gauge transformation parameterized by $\hat h ^{-1}$. This
leads to the following section
\begin{equation}
X=UX_d U^{-1}
\label{ins}
\end{equation}
and the gauge connection
\begin{equation}
A_z= -i (\partial U)U^{-1}, A_{\bar z}= -i (\bar \partial U)U^{-1}.
\label{ins1}
\end{equation}
It is easy to see that the pair $(X,A)$ forms a matrix string
instanton. Indeed the connection $A$ is both unitary and flat
while the section $X$  satisfies $\bar D_A X=0$ and the reality
condition $[X,\bar X]=0$.
We have thus found that matrix string instantons are
characterized by holomorphic sections of $Adj P$. 

One can also perform a multivalued gauge transformation by $U^{-1}$
which sends $X$ into the diagonal matrix $X_d$. Notice that
the corresponding gauge field vanishes altogether. As the eigenvalues of $X$
forming the diagonal
matrix $X_d$ are  solutions of an algebraic equation of degree $N$ we can
identify the instantons with a fibred space over the set of $N$-sheeted coverings of
the Riemann surface $\Sigma$\cite{don}. 
Moreover
the instanton is simply related to the diagonal matrix $X_d$ with the $\lambda^i$'s
along the diagonal.
The characteristic polynomial of $X_d$ can be expanded
\begin{equation}
\prod_{i=1}^N(\lambda-\lambda_i)=\lambda^N +a_1\lambda^{N-1}+\dots a_{N}
\end{equation}
where the coefficients $a_i$ are single-valued on $\Sigma$. One can reconstruct a matrix
$X_0$ with this characteristic polynomial\cite{tom}
\begin{equation}
X_0=\left (
\begin{array}{cccc}
-a_1&-a_2&\dots&-a_N\\
1&0&\dots&0\\
0&1&\dots&0\\
0&\dots&1&0\\
\end{array}
\right ).
\end{equation}
Having obtained  such a matrix one can use the relations (\ref{ins})
and (\ref{ins1}) to define the matrix string instanton. 

The analytic structure of $U$ is worth emphasizing. 
The eigenvalues of $X_0$ are $z$ dependent,
they  collide at branch points $B_i$. Some  eigenvalues collide
at such a point, the monodromy of the matrix  $U$
corresponds to the permutation $g_i$ of the coincident eigenvalues 
\begin{equation}
U\to U g_i
\end{equation}
The matrix $U$ is multivalued on $\Sigma$. Nevertheless the gauge connection (\ref{ins1}) is single-valued
implying that the $U(N)$ instantons are well-defined on $\Sigma$.

 One way of uniformizing
the behaviour of $U$ is to define the spectral cover $S$ of the Riemann surface
$\Sigma$.
Consider the set of solutions of the characteristic polynomial of $X$
\begin{equation}
\det (\lambda -X)=0
\end{equation}
This defines the spectral cover  of the multivalued instantons on $\Sigma$.
The spectral cover  is a $N$-covering of the Riemann surface $\Sigma$ ramified at the points
$B_i$ where some of the eigenvalues coincide. More precisely the
$N$ eigenvalues $\lambda_i$ define the inverse image of the
projection $\pi:S\to \Sigma$ sending $\lambda \to z(\lambda)$. The
$N$ sheets are defined by solving the equation $z(\lambda)=z$. On
each of the $N$ sheets there are holes corresponding to the
inverse images of the circles $a_{\pm}^i$.
Each sheet is in correspondence with a sphere with $h$ holes and
branch points connecting the different sheets. The branch points
 on the spectral cover are points where $dz/d\lambda$ vanishes.

 The matrix $U$ is extended to the spectral cover $S$ 
by considering that  $ U_{ij}$ connects the ith and jth
sheets.  
The permutation matrix $g_i$ is the monodromy matrix around the branch point
$B_i$ corresponding to the shuffling of the different sheets of the covering. 
The genus
of the spectral cover is given by
\begin{equation}
g_S=1+N^2(h-1).
\end{equation}
This can be derived 
 using the
Riemann-Hurwitz formula for generic  branched points  of order two.

The same analysis can be performed when marked points at $x_i$  are
added. The one form $X$ is now required to satisfy a boundary
condition in a neighbourhood of each marked point. One looks for
solutions with a simple pole at each marked point. One needs to introduce another Poincar\'e series
\begin{equation}
\theta[x,x_0]dz=\sum_{\gamma\in
\Gamma}adj(H_{\gamma}^{-1})d\ln\frac {\gamma (z)-x}{\gamma(z)-x_0}.
\end{equation}
The holomorphic solution becomes
\begin{equation}
X_0=\sum_{i=1}^hw_i(x_0)M_idz + \sum_{i=1}^v \theta[x_i,x_0]p_i dz
\end{equation}
and the residue condition 
\begin{equation}
\sum_{i=1}^h(adj(H_i^{-1})-1)M_i+\sum_{i=1}^v p_i=0.
\end{equation}

The matrix string instantons are obtained by diagonalization of $X_0$. 
The gauge connection is also   single-valued on $\Sigma$. 
One can similarly define the   spectral cover $S$ which uniformizes the behaviour of the matrix $U$. Its genus is
\begin{equation}
g_S=1+N^2(h-1)+v\frac{N(N-1)}{2}.
\label {genus}
\end{equation}
The marked points are lifted to $p=vN$ points on the spectral
cover.

 The genus formula is a consequence of the Riemann-Hurwitz relation
for an $N$-sheeted cover with
$N(N-1)(2h-2+v)$ branched points of order two. To obtain this
formula notice that the branch points are  obtained as
the common zeros of $P(\lambda,z)=\det (X-\lambda)$ and $
\partial_ {\lambda} P$. Factorizing $P=\prod_i(\lambda-\lambda_i)$
and $\partial_{\lambda} P=\prod_j(\lambda-\lambda'_j)$ where
the $\lambda'_j$'s are the zeros of the derivative of $P$, common
zeros are detected by the vanishing of the discriminant  $\Delta =\prod_{ij}(\lambda_i-\lambda'_j)$.
This corresponds to the zeros of $N(N-1)$ meromorphic one forms
with $v$ poles. Geometrically these are the $N(N-1)$
intersection points between the  curves $P=0$ and $\partial P=0$. Each one form
possesses $2h-2+v$ zeros leading to $N(N-1)(2h-2+v)$ branch
points. 

One can define a stratification of the moduli space according to the genus
of the spectral cover \cite{nekra}. As just seen the genus of the 
spectral cover depends on the number of branch points via the Riemann-Hurwitz formula.
The branch points are the zeros of the discriminant $\Delta$ viewed as a $N(N-1)$ differential on $\Sigma$.
Denote by $p^d_i$ the eigenvalue matrix obtained by diagonalization of the residue
of $X_0$ around the pole $x_i$. This matrix possesses $k_i^m$ eigenvalues of order $m$.
The order of the pole of $\Delta$ at $x_i$ is 
\begin{equation}
o_i=N^2-\sum_{m=1}^N m^2 k^m_i.
\end{equation}
The number of branch points is then
\begin{equation}
2N(N-1)(h-1)+\sum_{i=1}^v o_i
\end{equation}
leading to the genus of the spectral cover
\begin{equation}
g_S=1+N^2(h-1)+\frac{\sum_{i=1}^v o_i}{2}.
\end{equation}
In the generic case where all the eigenvalues are different one gets $k_i^1=N$ and the previous formula
(\ref{genus}). This defines the generic stratum of the moduli space. Other lower dimensional strata are 
obtained when there are multiple eigenvalues. The generic stratum is an open dense subset of the moduli space
whose boundary comprises the other strata.

We are now in position to compute the dimension of the moduli
space of instantons. This dimension is given by the number of
independent matrices $H_i$, $M_i$ and $p_i$ modulo a  residual global
symmetry.
Once the holomorphic field $X_0$ is defined in terms of the
parameters $(H_i,M_i,p_i)$ the gauge field is determined by the
 matrix $U$. 
The  adjoint action of the residual symmetry $Sl(N)$ on $(H_i,M_i,p_i)$ is
deduced from
 $X_0\to adj(V)X_0,\  h\to Vh$.  This 
reduces the real dimension of the moduli space by $2(N^2-1)$. The
numbers of parameters $(H_i,M_i,p_i)$ modulo the residue condition is $4N^2h+2vN^2-2N^2$.
Combining with the residual symmetry we get $\hbox{dim} {\cal M}= 2N^2(2h+v-2)+2$.
Using the genus formula we obtain
\begin{equation}
\hbox{dim}_{C}{\cal M}=2g_S-1+p
\end{equation}
This dimension coincides with the dimension of the space of Higgs
bundles\cite{don} which precisely correspond to the complex vector
bundles $Adj P$  with a section of $K_{\Sigma}\otimes Adj P$. This is not unexpected
as we started with such a section $X_0$ to define the instantons.

Let us now consider the lower strata of the moduli space. It is also possible to count the number of moduli.
The main difference with the generic stratum comes from the number of moduli
preserving  coinciding eigenvalues. It is easy to see that changes of basis
in $Gl(N)/\prod_{m=1}^N Gl(m)^{k_i^m}$ preserve the eigenvalues of $p_i$. This provides
 $\sum_{i=1}^v\sum_{m=1}^N k^m_i$ complex
 eigenvalues
and $\sum_{i=1}^v o_i$ parameters for the changes of basis.
Now the number of marked points  on the spectral cover is 
\begin{equation}
p=\sum_{i=1}^v\sum_{m=1}^N k_i^m
\end{equation}
It follows  that the number of moduli is still given by (\ref{mod}).
We can get rid of one dimension of the moduli space by
factorizing the complex rescalings of $X_0$. This leads to a
dimension   $ \hbox{dim}_{C}{\cal M}=2g_S-2+p$ for the moduli space of instantons. 
This formula has a simple geometric interpretation.

The link with the spectral cover is provided by the existence of a natural one
form $\pi^* X_d$, the pull-back of the one form $X_d$ under the
projection $\pi: S\to \Sigma$ defined by $\lambda \to z(\lambda)$.
 Locally the pull-back is defined
by $\lambda dz$ where $\lambda$ defines the coordinates on the
spectral cover.
This one-form on the spectral cover possesses a pole at all the
inverse images of the poles of $X_d$. The number of zeros $I_i$ of $\pi^*
X_d$ is then $2g_s-2+p$. On the generic stratum they are of two sorts, there are the $N(N-1)(2h-2+v)$
branch points $B_i$ and the $N(2h-2+v)$ inverse images of the zeros
$Z_i$ of each 
of the $\lambda_i$'s.  As we vary the parameters of the
instantons the zeros $I_i$  move.

 Consider a spectral cover
$S_0$ with zeros $I_i^0$ and a projection $\pi_0: S_0\to \Sigma$
corresponding to an instanton $X_0$. This spectral cover
is defined by the values $(H_0,M_0,p_0)$ of the instanton parameters.
For different values of these parameters one obtains a family of
spectral covers parameterized by $2g_S-2+p$ coordinates $t^i$ on
the instanton moduli space. Each spectral cover is endowed with a
projection $\pi_t$ depending on the moduli $t$. Locally
 one can define a map $\delta=\pi^{-1}_t\pi_0 :S_0\to
S_t$ as $\lambda_0^i\to \lambda^i(z_0(\lambda_0^i))$  
encoding the deformations of $S_0$ due to the variations of the
moduli $t$. This defines a $2g_S-2+p$ family of spectral covers $S_t$.
The family of forms $\pi_t^* X_d$ on $S_t$ vanish on the zeros
$I_i(t)$. One can use the coordinates of the zeros $I_i(t)$ to parameterize the moduli space. 

More precisely one can reconstruct the $U(N)$ instantons from the knowledge of the
spectral cover and a line bundle ${\Omega_P}$ over $S$. Indeed let us consider the gauge where the gauge connection vanishes and 
the instanton is reduced to a diagonal  matrix of one forms subject to monodromies around the branch points.
Similarly this matrix can be viewed as arising from the  single form $\pi^* X_d$ defined on $S$ under the push-down map
$\pi_*$ sending $\pi^* X_d=\lambda dz$ on $S$ to $\lambda^i dz$ on $\Sigma$. The one form $\pi^* X_d$ is a section of
 the line bundle $\Omega_P$ of one forms $s$ whose divisors satisfy $(s)+P\ge 0$. Conversely let 
consider  the line bundle $\Omega_P$ and its push-down 
$\pi_* {\Omega_P}=\bigoplus_{i=1}^N V_i$
the direct sum of one dimensional vector spaces. Choose a section $s$ of ${\Omega_P}$ and its push-down $\pi_* s$.
This defines a diagonal matrix of one forms on $\Sigma$ from which one can obtain the characteristic polynomial and then an
instanton. One can write the divisor $(s)=I-P$ for an effective divisor 
$I$ of degree $2g_S-2+p$ where $s$ vanishes. This divisor of zeros parameterizes the moduli space of instantons.

 It is more
convenient to factor out a trivial factor from the moduli space
corresponding to the position of one zero.
The reduced
moduli space has dimension
\begin{equation}
\hbox{dim}_{C}{\cal M}_R=2g_S-3+p.
\end{equation}
  The dimension of the
reduced moduli space coincides with the result of a calculation
using an index theorem\cite{nesti3}.
We have thus confirmed  this indirect argument by an explicit construction.
\section{Mandelstam Diagrams and the Moduli Space of Instantons}

We shall be  concerned  with the matrix string setting,
i.e. the Riemann surface $\Sigma$ is obtained by a conformal map
of the cylinders with $(v-2)$ marked points to a sphere with $v$
marked points. In this case the Poincar\'e series parameterizing
the instantons are simply 
\begin{equation}
X_0= \sum_{i=1}^v p_i  \frac{1}{z-x_i}
\end{equation}
where $\sum_{i=1}^v p_i =0$. Diagonalizing this matrix leads to the spectral cover $S$.
In this case the genus of the spectral cover is simply
\begin{equation}
g_S=1-N^2+\frac{N(N-1)v}{2}
\end{equation}
on the generic stratum.
Notice that on the generic stratum 
$g_S\ge 0$ with equality only when $p=2N+2$. The dimension of the moduli space is simply
given by the number of matrices $p_i$ modulo the residual global symmetries and the momentum conservation
 condition $\sum_{i=1}^v  p_i=0$. For instance one can describe the genus zero scattering of six external states with a $(N=2)$
 covering of the cylinder with one marked point.

 There is an alternative
description of the spectral cover which will be particularly fruitful.
 Let us
recall a few useful facts about Mandelstam diagrams \cite{dhokerphong,mandel}.  
Mandelstam diagrams represent the propagation of circular strings
joining at interaction points. For strings of length $2\pi$
the radii of the cylinders are identified with the momenta $p^+$
in the light cone frame.  A defining property of the
Mandelstam diagram of a Riemann surface  is the existence of a  time 
axis $\tau$  and of periodic angular coordinates $\sigma$   on each of the cylinders.
The complex variable defined by $w=\tau +i\sigma$ parameterizes
each of the cylinders.

A Mandelstam diagram is uniquely specified by its interaction
points  where $dw$ vanishes, $g_S$ internal momenta and $g_S$ internal shifts.
 On the whole this provides $3g_s-3+p$ complex
coordinates forming a cover of the moduli space of marked
Riemann surfaces \cite{gidd}.
 Finding these parameters will allow us to construct the Mandelstam diagram
associated to the interaction of strings on the spectral cover.

\begin{figure}
\epsfxsize=12.cm
$$\epsfbox{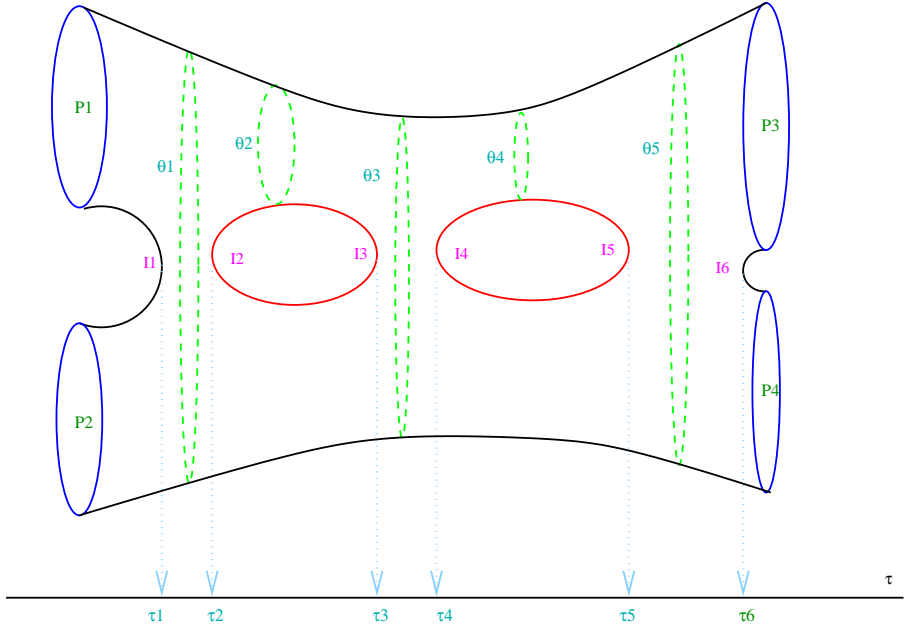}$$
\caption{The string diagram of the spectral cover with interaction points $I_i$,  $i=1..6$.
The interaction times $\tau_i$ and the angles $\theta_i$ define the location of the interaction points. The external states correspond to the marked points $P_i$, $i=1..4$}
\end{figure}

The  construction of a Mandelstam diagram has to be adapted
to the case of the spectral cover.   Let us use the function
\begin{equation}
\tilde w(\lambda)=\int^{\lambda} \pi^* X_d
\end{equation}
defined up to the periods of $\pi^* X_d$. Consider
now the real part
\begin{equation}
\tilde \tau= \mbox{Re}(\tilde w)
\end{equation}
up to the real parts of the periods.
This real function has critical points at the zeros $I_i$ with an
index $-1$\footnote{This gives a contribution $2-2g_S-p$ to the
Euler characteristic.}. At
the poles one sees that $\tilde \tau \to \pm \infty$. The function
$\tilde \tau$ is a  Morse function  allowing
to stratify the spectral cover  according to its level sets. The image of the
spectral cover under $\tilde w$  is interpreted as a string diagram with
interaction points at the zeros $I_i$ corresponding to the
splitting or joining of two strings. Topologically this string
diagram is defined  by $2g_S+p-3$ complex coordinates
corresponding to the interaction points
 The position of the
interaction points
is specified by the  interaction times  and the twist angles
along $2g_s-3+p$ branches of the string diagram (fig. 1).

The spectral cover can be given a metric using the string diagram.
Let us define the flat metric $d\tilde wd\overline{\tilde w}$  on the
spectral cover. This is the induced metric from the flat metric
on the string diagram. On the spectral cover we obtain
\begin{equation}
g_{\lambda\bar\lambda}=\vert \lambda \frac{dz}{d\lambda}\vert^2
\end{equation}
Due to the holomorphy of the zweilbein $w_{\lambda}=\lambda dz/d\lambda$
the metric is almost flat apart from the zeros and the poles
creating a delta function singularity in the curvature
\begin{equation}
R(Q)=2\pi \sum_{i=1}^{2g_S-2+p}\delta(Q-I_i)-2\pi\sum_{i=1}^p \delta(Q-P_i)
\end{equation}
Of course the integral of the curvature gives the Euler
characteristic.
 This is consistent with the string interpretation
of the string diagram where the strings are flat cylinders
joining at the interaction points.
The string diagram defines the dynamics of the string
interactions, i.e. a contact interaction at the
interaction points. 
The kinematics of the strings is not yet specified as we have not
explicitly described the momenta of the incoming and
outgoing strings. This is provided by the addition of Wilson
lines on the spectral cover.    
   
Up to now we have not taken into account the discrete $Z_N$ twists
along homology cycles which can be turned on in the construction
of flat bundles.
 This corresponds to
adding different twisted sectors in the quotient $ U(1)/Z_N$.
The flat vector bundle with the gauge connection $A$ can be viewed as the quotient
$(\hat \Sigma \times su(N)\times S_1)/\sim$ where $\hat \Sigma$ is the universal cover
of $\Sigma$, $su(N)$ the Lie algebra of $SU(N)$ and $S_1$ the unit circle. The equivalence relation is
$(x,u,\theta)\sim (\gamma (x), adj(\gamma)u,\theta)$ where $\gamma$ is a loop on $\Sigma$ acting
on $\hat \Sigma$ as covering transformation and $adj(\gamma)$ is the monodromy in the adjoint representation.
Notice that the $S_1$ factor is not involved in the quotient. One can define a twisted version of the
equivalence relation $(x,u,\theta)\sim (\gamma(x),adj(\gamma)u,\theta +\int_{\pi^{-1}(\gamma)}a_S)$
obtained by integrating  a $U(1)/Z_N$ connection $a_S$ on $S$ along the inverse image of $\gamma$.
This twisting modifies the $U(1)$ part of the gauge connection. It does not change the field $X$.

Let us consider a $U(1)$ bundle over the spectral cover $S$ that we endow with a
flat connection $a_S$. The building of such a flat vector bundle
is characterized by maps in $\mbox{Hom}(\pi_1(S),U(1))$. This corresponds to 
having different winding numbers along the non-trivial loops of $S$.
The flat vector bundle is modeled on the bundle $(\tilde S\times U(1))/\sim$ where
$\sim$ is the equivalence relation $(x,\theta)\sim (\gamma (x), \theta+ 2\pi d_{\gamma})$,
$\tilde S$ is the universal cover of $S$ and $\theta$ is an angle. The winding number
 $d_{\gamma}$ depends on the 
loop $\gamma$ which acts on $x$ as a covering transformation.  
We can now easily incorporate the $Z_N$ orbifold    characterized by
maps in $\mbox{Hom}(\pi_1(S),Z_N)$. For a given loop $\gamma$ in
$\pi_1 (S)$ one can define a twisted action with twist $2\pi k/N$
by $\theta \to \theta +2\pi k/N,\ k\in \ [0,N-1]$. The range of $k$ is extended to the whole
integers by taking into account the non-trivial winding numbers. As the fundamental group of
the spectral cover is of dimension $2g_S$  this gives
$N^{2g_S}$ sectors. Physically these discrete 
transformations are discrete Wilson lines in the $U(1)$ part of
the gauge theory.

One can always construct such a flat $U(1)$ gauge field $a_S$. Put
$a_S=b+\bar b$
where $b$ is a holomorphic one form $b=\sum_{i=1}^{g_S}u_i \omega_i$ and  the  one forms $\omega_i$ form
a basis of holomorphic Abelian differentials. Now we prescribe the Wilson lines
\begin{equation}
\int_{a_i} a_S= \frac{2\pi k_i}{N}, \ \int_{b_i}a_S=\frac{2\pi k'_i}{N}.
\end{equation}
The complex coefficients $u_i$ are always uniquely defined as the period matrix
$\Omega_{ij}=\int_{b_j}\omega_i$ has an invertible imaginary part. 
One  obtains the  push-down of $a_S$ to $\Sigma$ as  $\pi_* a_S=(a_d)$, i.e.  a diagonal matrix of one forms
$b(\lambda^i)d\lambda^i+\bar b(\bar \lambda^i)d\bar\lambda^i$ where $a_S=b(\lambda)d\lambda+\bar b(\bar\lambda
)d\bar \lambda$.
This matrix possesses monodromies around the branch points corresponding to the
exchange of the different sheets. The $U(1)$ gauge field 
\begin{equation}
a_{\Sigma}=\mbox {tr} (\pi_* a_S)I_{N\times N}
\end{equation}
defines a $U(1)$ flat gauge connection on $\Sigma$. This connection is diagonally embedded
in $U(N)$. The total gauge connection is the sum of $a_{\Sigma}$ and the connection 
(\ref{ins1}).

As will be recalled in the next section the semi-classical
evaluation of the matrix string path integral involves a residual
$U(1)$ gauge field $a$ \cite{nesti2} living on the spectral cover.
The other fields  correspond to strings degrees of freedom in
the Green-Schwarz formulation. Considering the Hamiltonian
picture of this theory on the spectral cover the physical states are
identified with light
cone string states carrying  some  information coming  from the
quantization of the extra $U(1)$ factor.

The discrete Wilson lines  modify the
behaviour of the string states propagating along the string
diagram of the spectral cover. 
 Given a   state
$\vert \alpha>$ located
at a non-trivial $a_i$ loop
on the string diagram (fig. 1)
and going along the cycle $a_i$ transforms the string state $\vert
\alpha >$ into $e^{2\pi ik_i/N}\vert \alpha >$. Normalizing the
string length to $2\pi$ we find that the state $\vert \alpha>$ acquires
a momentum $k_i/N$. Similarly the same state transported along
the cycle $b_i$ picks up a phase $e^{2\pi i k'_i/N}$ corresponding
to a shift  of the origin of the string by $ k'_i/N$.

The marked points correspond to the location of the external vertex operators.
One can open up a small disc around each marked point to obtain a puncture.
The punctures lead to  $(p-1)$ extra cycles. 
The $(p-1)$ cycles around the punctures give  quantized  external
momenta $k''_i$. This is achieved by modifying the $U(1)$ flat connection demanding
that
\begin{equation}
\int_{\gamma_i}a_S= \frac{2\pi  k''_i}{N}
\end{equation}
where $\gamma_i$ are the $p$ cycles around the punctures.
One can always find the gauge connection as
 $a_S=b+\bar b$ where $b=\sum_{i=1}^{g_S}u_i \omega_i +\sum_{i=1}^{p}v_i
\omega_{P_0 P_i}$ and
 $\omega_{P_0P_i}$ is a meromorphic one form with poles at $P_i$ and $P_0$ with residues 
one and minus one respectively. The periods of these one forms $\omega_{P_0P_i}$ are imaginary along the $a$ and $b$ cycles.
Requiring that $b$ is holomorphic at $P_0$ implies the momentum conservation $\sum_{i=1}^p k''_i=0$.

\begin{figure}
\epsfxsize=12.cm
$$\epsfbox{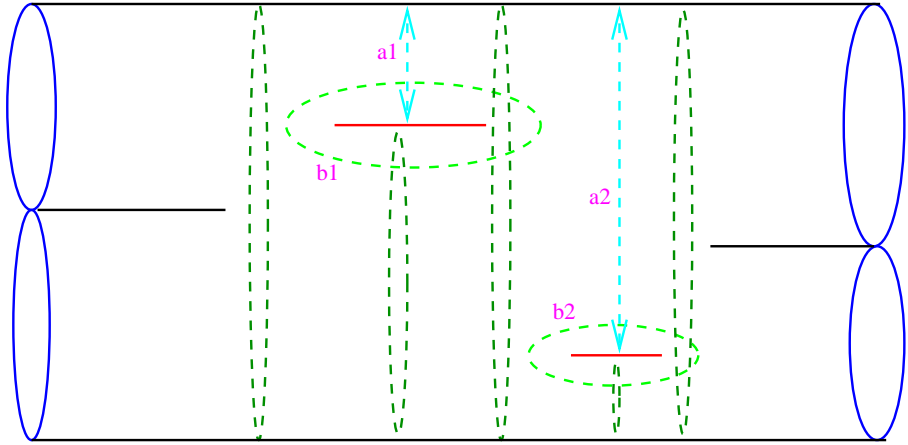}$$
\caption{The equivalent Mandelstam diagram to the spectral cover where the momenta along 
the cycles $a_i$, $i=1,2$ and the twist angles along the cycles $b_i$,  $i=1,2$ are quantized}
\end{figure}

The picture emerging from the description of the spectral cover
 fits with what is expected from  light cone string
theory\cite {wolpert}.
First of all the continuous part of the construction leads to the
spectral cover and its interpretation as a string diagram for
closed strings. The parameterization of the spectral cover in
terms of the coordinates of the interaction points specifies the
interaction times and $2g_s-3+p$ shifts. The discrete part
of the construction, i.e. the assignment of $Z_N$ elements  to
the $2g_S+p-1$ non-trivial cycles is equivalent to prescribing
the internal and the external momenta in a quantized fashion. The
strings are also twisted along the $b$ cycles with discrete shifts $s_a,\ a=1\dots g_S$.

With this information we can construct the Mandelstam diagram of
the spectral cover $S$.  One can assign  to all the
 $p$ external cylinders and the $(3g_S-3+p)$
internal cylinders a radius equal to the quantized momentum $\alpha$ as determined by momentum conservation. 
This requires to define new  coordinates on each of the cylinders
by rescaling  $\tilde w$ by $\alpha/L$ where $L$ is the radius measured using the
coordinates $\tilde w$. The radius  of each of the
cylinders is now $\alpha$. Globally one patches up the new coordinates $ w$ on each of the
cylinders by imposing that going around the $b$ cycles the imaginary part of $ w$ is shifted by
$s_a$. The coordinates
\begin{equation}
 w=\tau+i\sigma
\end{equation}
 built this
way are defined globally and describe the Mandelstam diagram.
The coordinates of the interaction points are now $
w_i=\tau_i+i\sigma_i,\ i=1\dots 2g_S-3+p$.

  The Mandelstam diagram of $S$  has been built by joining the $p$ external cylinders
and  the $3g_S-3+p$ internal cylinders. The relative orientation of
 the flat cylinders is prescribed by the different
shifts $\sigma_i\ i=1\dots (2g_S-3+p)$ and $s_{a}\ a=1\dots g_S$. The length
of the cylinders is given by the interaction times $\tau_i$. In another parameterization
 which will be useful when
 identifying the measure over the instanton moduli space one can utilize
  the twisting angles
\begin{equation}
\theta_i=2\pi \frac{\sigma_i}{\alpha_i},\ \eta_a=2\pi \frac{s_a}{\alpha_a}
\label{twist}
\end{equation}
where the $\alpha_i$'s and the $\alpha_a$'s are  the quantized momenta along the
corresponding internal cylinders.
We have thus constructed a Mandelstam diagram from the coordinates of the interactions points $I_i$
 and the quantized external and internal momenta.

Deformations of the Mandelstam diagram by varying the positions
of the interaction points, the  shifts and the momenta correspond to
changing the parameters of the instantons.   
 On the whole we have identified the moduli space of instantons
as  a
discretization of the moduli space of marked Riemann surfaces. This result had already be
obtained in \cite{nesti3} using independent arguments. Here we
have shown how this decomposition is intimately linked to  the moduli space of matrix
string instantons. In the next section we will study 
 the dynamics of the string interactions more carefully and come back
to the essential  $U(1)$ factor  in the
semi-classical approximation.

\section{ \bf The Scattering Amplitudes}

\vskip 1 cm
The path integrals of matrix models can be evaluated in a  semi-classical fashion
 in the small string coupling limit\cite{nesti2}.
The objects of interest are the scattering amplitudes defined
by the path integral
\begin{equation}
A(1,....,n)=\int {\cal D}A {\cal D} X{\cal D}\Theta \phi(0) \phi(\infty) V_1...V_{v-2} e^{-S(X,A,\Theta)}.
\end{equation}
 The vertex
operators are the analogue of the light cone string vertex
operators. We also have inserted a wave function at both ends of
the cylinder. These wave functions depend on the boundary values
of the fields on the external circles.  The link between the vertex operators and the
wave functions is the usual one, i.e. obtained by integration over 
 all the field configurations on a small disc with a vertex
operator inserted at its centre. The vertex operators  only involve the
diagonal part of the fields in a suitable gauge.
The string interpretation of the scattering amplitudes involves
the semi-classical evaluation and the lifting of the action to
the spectral cover.

Let us first deal with the action in the semi-classical regime \cite{nesti2}.
We  expand the action around the instantons.
The instantonic configuration $(A,X)$ is taken in the gauge where
the bosonic field $X_d$ is diagonal. This amounts to applying a
multivalued gauge transformation $U^{-1}$ to all the fields. In
particular the diagonal fluctuations $x_d^i$ 
are multivalued due to the non-trivial monodromies at the branch
points.

 The path integral measure needs to be defined
by fixing a gauge and introducing Faddeev-Popov ghosts. The gauge
is fixed by
\begin{equation}
G_{w\bar w}=\partial a_{\bar w}+\bar \partial a_{w}+ ig^2([X_d,\bar x]+[\bar X_d,x]).
\end{equation}
Capital letters are only used for the background instantons.
The Fadeev-Popov action is
\begin{equation}
S_{FP}=-\frac {1}{2\pi g^2 }\hbox {tr} (\int d^2 w \bar c \frac {\delta
G}{\delta \epsilon} c)
\end{equation}
where $\epsilon$ is the gauge variation parameter.
The gauge fixing is simply
\begin{equation}
S_{GF}=\frac{1 }{4\pi g^2}\hbox {tr} (\int d^2 w G^2).
\end{equation}
The fields are defined by an appropriate rescaling. The diagonal
fields $a^d,c^d$ are rescaled by a factor of $g$.  
\begin{equation}
A_w=ga_w^d+a_w^{nd},\ X=X_d + x_d+ \frac {x_{nd}}{g}, \
X^I=x^I_d+\frac{ x^I_{nd}}{g}, \ \Theta=\theta_d +\frac
{\theta_{nd}}{\sqrt g}
\end{equation}
where $I=3..8$ are the directions transverse to the instanton.
The ghosts are likewise
\begin{equation}
c=gc_d+\sqrt g c_{nd}
\end{equation}
and similarly for $\bar c$.
The Euclidean action becomes
\begin{equation}
S=S_d+S_{nd}+O(\frac {1}{\sqrt g})
\end{equation}
The non-diagonal action $S_{nd}$ is quadratic and  can be
integrated out giving unity by supersymmetry.
The higher order terms are all suppressed by the gauge coupling constant.
In the small string coupling limit we are left with a purely diagonal
path integral.
The normalization of the diagonal fields has been chosen in such
a way that the quadratic action is independent of the string
coupling constant.

The  evaluation of the path integral is not 
trivial due to the
multivalued nature of the diagonal fields.
To obtain single-valued fields one needs to lift the fields  to the spectral
cover. The spectral cover is
defined by considering the instanton on the Riemann sphere
$\Sigma$ and pulling it back to the cylinder $C$ via the
conformal mapping $f:C\to  
\Sigma$.
One gets $X_C= f^* X_{ \Sigma}$. The spectral cover of the
cylinder is then obtain from the eigenvalue equation $\det
(\lambda -X_C)=0$.
This allows use to use the previous results and lift the fields
from the Riemann sphere to the spectral cover.

The diagonal vectors  $x_{dj},\ j=1\dots N$ define  a
section $x$ of a line bundle on the spectral cover $S$ by the rule
$x(\lambda^i(z))=x^{i}(z)$ where $\lambda^i(z)$ is the ith eigenvalue
of $X_d$ at $z$.
Similarly the gauge field is lifted  to the spectral cover
leading to a one form $a^i(z)= a(\lambda^i) d\lambda^i/d\omega $.
This leaves us with an action defined on the
spectral cover $S$ for the bosonic fields
\begin{equation}
\frac {1}{\pi}(\int d^2\lambda g^{\lambda\bar \lambda}\partial_{ \lambda}a_{\bar \lambda}
\partial_ {\bar \lambda}a_{\lambda}+ \int d^2\lambda \partial_\lambda x^I_d\partial_{\bar \lambda}
x^I_d)
\end{equation}
The integration is over $S$. Notice that the metric vanishes at the interaction points.
At these points the inverse metric is ill defined and the
semi-classical approximation breaks down. 
The action for the ghost fermions $c,\bar c$ is easily derived
\begin{equation}
\frac {1}{\pi} \int d^2\lambda  \partial \bar
c\bar \partial c
\end{equation}
The ghost fields are anticommuting scalars on the  spectral cover.

We can now turn to the space time fermions.
As defined in the path integral over  the cylinder these fields are
anticommuting scalars on the world sheet. They can be promoted to
world sheet fermions on the spectral cover. This requires the use
of the  one form $\omega_{\lambda}$ and its square root.
The square root of the one form $\omega_{\lambda}$ requires the choice of a
spin structure on the spectral cover. This amounts to choosing  the square root along
 non-trivial cycles. The spin structure
is specified by $2g_S$ signs corresponding to such a choice.
Once a spin structure has been chosen it is called even if the product of the $2g_S$ signs
 is even and odd in the opposite case.  This endows the
spectral cover with the structure of a super-Riemann surface.
To maintain the physical interpretation of the space-time fermions  we impose that
 the square root of $\omega_{\lambda}$ picks up a minus
sign on all the possible
cycles of the
 string diagram. The spin structure is then even.

This allows to transform the space time fermions into world sheet
fermions with values in the spin representations of $SO(8)$
\begin{equation}
\psi^{\alpha}=\sqrt {\omega_{\lambda}}\theta^{\alpha},\ \bar \psi^{\dot\alpha}=
\sqrt {\bar \omega_{\lambda}}\theta^{\dot\alpha}.
\end{equation}
The action for the fermions $\psi,\bar \psi$ is lifted to the spectral
cover to give the usual Dirac action
\begin{equation}
\frac{i}{\pi}\int d^2 \lambda (\bar\psi^{\dot\alpha} \partial \bar \psi_{
\dot \alpha}+ \psi^{\alpha}\bar \partial \psi_{ \alpha}).
\end{equation}
Notice that the choice of the spin structure implies that there
are cuts ending at the interaction points and the marked points.

On the whole we find  that the end-result of the semi-classical
approximation is to produce a theory defined on the spectral
cover of the matrix string instantons. Once the spectral cover is
specified the theory is given by the light-cone Green-Schwarz
action for space-time bosons and fermions. The space time fermions are
also world sheet fermions on the spectral cover. This theory is
coupled to a  $U(1)$ theory.

Before completing the analysis of the path integral it is noteworthy 
to consider the Hamiltonian formulation of the theory. This is most conveniently
achieved by considering the flat coordinates $\tilde  w$ and the time
axis $\tilde \tau$ on the string diagrams. Using these coordinates the
actions for the scalars, fermions and gauge fields reduce to
 free field theories defined on the flat cylinders forming the
string diagrams. In the Hamiltonian picture there are
Green-Schwarz string states $\vert GS>$ propagating along each of
these cylinders and interacting at the interaction points.
Because of the presence of non-trivial  $U(1)$ backgrounds $a_S$ the intermediate states $\vert\alpha>$ are also
characterized by  fractional momenta $\alpha$ on all the
internal and external cylinders.
 There are also fractional  shifts $s_a$
for $g_S$ internal  cylinders. The Hilbert space of  the theory is 
\begin{equation}
{\cal H}={\cal H}_{GS}\otimes {\cal R}
\end{equation}
where ${\cal H}_{GS}$ is the Hilbert space of
the GS states $\vert GS>$ and ${\cal R}$ is the Hilbert space
deduced from the quantization of the $U(1)$ theory.

Consider one of the cylinders on the spectral cover, we normalize
its radius to $L$. The corresponding $U(1)$ action
is then
\begin{equation}
\frac{1}{2\pi g^2}\int_0^{2\pi L}d\tilde \tau d\tilde \sigma (\partial_0 a_1)^2
\end{equation}
in the gauge $a_0=0$. 
In this gauge it is convenient to introduce the Wilson line
\begin{equation}
g=e^{2\pi i L a_1}
\end{equation}
at a given time $\tilde \tau$ for $\tilde \sigma$ independent gauge fields.
In terms of this Wilson line the action reads
\begin{equation}
\frac{1}{4\pi^2 Lg^2}\int d\tilde \tau \vert \dot g\vert ^2
\end{equation}
and the Hamiltonian
\begin{equation}
H=4\pi^2 L g^2\vert p\vert^2.
\end{equation}
This is the Hamiltonian of a free particle on a circle.
The motion of this free particle depends on the different
types of boundary conditions, i.e. on the discrete Wilson lines.
In the untwisted sector the circle is identical to
$R/2\pi  Z$. In the twisted sectors with a Wilson line $k/N$ 
the eigenvectors pick up a phase $e^{2\pi ik/N}$. In all cases denoting by
$\theta$ the angular variable  the momentum reads $i\partial/\partial\theta$
with eigenvectors $e^{2\pi i {(n+k/N)} \theta}$.
The states $\vert n,k>$ associated to these eigenvectors form a
basis of the $U(1)$ Hilbert space on the cylinder.
There is one more subtlety due to the possible shifts $s_i$ along the
$b_i$ cycles bording the cylinder. The Hilbert space ${\cal R}$ is
spanned by all the $\vert n,k, s_i>$ which are eigenstates of $H$
on the different cylinders and having a monodromy $e^{2\pi i s_i}$ along
the $b_i$ cycles. 

Let us come back to the energy associated to these eigenstates.
It simply reads
\begin{equation}
H\vert n,k,s_i>=4\pi^2 Lg^2(n+k/N)^2\vert n,k,s_i>.
\end{equation}
Combined with the eigenstates of the GS Hamiltonian $H_0$ acting
on light cone GS states we obtain eigenstates spanning the
Hilbert space ${\cal H}$.
The total Hamiltonian reads $H_T=H+H_0$.
The propagation of the eigenstates from one end of the cylinder to
the other end gives the evolution operator in Euclidean time
\begin{equation}
e^{-\tau (4\pi^2Lg^2(n+k/N)^2+(p^I)^2+m^2)}
\end{equation}
where $m$ is the mass of the GS state and $p^I$ its transverse momentum.
Notice that the corrections to the GS eigenvalues are
non-perturbative and of order $1/g_s^2 $. In the limit $g_s\to
0$ the only non-vanishing terms satisfy $n=0$, i.e. the
contributions from the winding modes decouple. Moreover the non-perturbative
corrections to the string scattering amplitude vanish altogether
provided $g_s^2N^2/L\to \infty$. For the long strings where $L\propto N$ the most stringent
condition is then
\begin{equation}
g_s^2  N\to \infty.
\end{equation}
In the finite $g_s$ and $N$ cases the string amplitude is
corrected by non-perturbative contributions coming from the creation
of D0 branes.

To study the string scattering amplitude we change coordinates
and use the $ w$ coordinates. In these coordinates the
radius of each cylinder is equal to the momentum $\alpha=k/N$.
The light cone Hamiltonian whose bosonic part is  
\begin{equation}
\frac{1}{2\pi}\int_0^{2\pi \alpha}
d\sigma((\frac {\partial x_d^I}{\partial\tau})^2
+(\frac{\partial x_d^I}{\partial \sigma})^2)
\end{equation}
has eigenvalues $2p_-$ where
\begin{equation}
p_-=\frac{(p^I)^2+m^2}{2\alpha}.
\end{equation}
The evolution operator is now
\begin{equation}
e^{-2ip_- \tau}
\end{equation}
where we have performed a Wick rotation on the world sheet. This
evolution operator coincides with the light cone string field
evolution operator. More precisely the evolution
operator is  associated to the
 Hamiltonian\cite{gidd}
\begin{equation}
K=i\alpha \frac{\partial}{\partial \tau} -H_0.
\end{equation}
This Hamiltonian is easily identified with the light cone string field
Hamiltonian 
with the canonical quantization rule
$[x^{\pm},p^{\mp}]=i$ yielding $p^- =-i
\partial/\partial x^-$. We obtain the following 
\begin{equation}
x^+=2\tau.
\end{equation}
Similarly with $p^+\equiv \alpha=-i\partial/\partial x^-$
we find that the $x^-$ coordinate is periodic, the radius of the circle
parameterized by $x^-$ is
\begin{equation}
R_-=N.
\end{equation}
This completes the identification of the small string coupling regime
of matrix string theory with light cone string theory where
the  radius $R_-$ goes to infinity with $N$.
In conclusion we have seen that the large $N$ limit is necessary
in order to decouple non-perturbative effects and decompactify
the $x^-$ direction.

We can now make  the above picture global.
To do so we use the global coordinates $ w$ patched up to
construct the Mandelstam diagram.
 By inserting a complete set of observables
at the  internal ends of the cylinders we build the scattering amplitude
by propagating the states with the $e^{-2i\tau_ip_-}$ evolution
operator on each internal cylinder.
We still have to specify the interaction vertices.
This necessitates to study  the ghost action and its zero modes.

\section{ Picture Changing Operators and Supermoduli}

The usual rules of
string perturbation theory prescribe the inclusion of picture
changing operators at the string interaction points\cite{mandel}. In the
RNS string context the picture changing operators appear after
integrating over the supermoduli\cite{verlin2,verlin3}. To do so one deforms the
super-Riemann surface structure by an anticommuting $-1/2$ tensor
$ v$
and fix the two-dimensional gravitino field $\bar \partial 
v$ to be equal to a linear combination of the super-moduli.
The deformation of the super-Riemann surface induces a
supersymmetry transformation of the bosons and fermions. This
leads to a direct coupling of the gravitino to the supersymmetric current.
The integration over the components of the gravitino yields an
insertion of the supersymmetric current at all the interaction
points and hence the picture changing operators.

In this section we will repeat parts of this construction. The
deformation of the super-Riemann surface will lead to the
inclusion of the picture changing operators at the interaction
points as already discussed in the DVV paper. 

In the following we shall need to  integrate functionally over various types of tensors on $S$. We
  define it by considering the Hilbert space of square integrable $n$ tensors defined by the scalar product
\begin{equation}
(\phi_n,\psi_n)=\int d^2 \lambda (g_{\lambda\bar \lambda})^{1-n}\bar \phi_n\psi_n.
\end{equation}
Imposing that the integral is well-defined in the vicinity of the
interaction points implies that for meromorphic tensors such that $\phi_n \sim 1/\lambda^p$ one should verify
 that $p<2-n$. Functions can have at most a pole at the
interaction points. The spinors $n=1/2$ can also develop a pole. The one-forms have to be holomorphic while
the quadratic differentials have to vanish. The same analysis
in the vicinity of a marked point where the metric admits a pole implies that $p<n$. Thus functions
 have to vanish at the marked points  while tensors of higher degrees can either be constant for $n=1$ or have
 a pole
$n\ge 2$.

Let us now study the  normalizable ghost zero modes. 
The gauge  fields $(a_{\lambda},a_{\bar \lambda})$ have been lifted to the spectral cover where they become
 single-valued. This entails that the only normalizable zero modes
 of the gauge fields are the $g_S$ holomorphic differentials. This is different for the ghosts.
 Indeed there are zero modes
of the fermionic ghosts which are normalizable though not single-valued on
the spectral cover. These zero modes are solutions of
\begin{equation}
\bar \partial c=0
\end{equation}
corresponding to meromorphic $1/2$
tensors $\psi_{1/2}$ on the spectral cover under
the world sheet
 boson-fermion correspondence sending $\psi_{1/2}\to \omega_{\lambda}^{-1/2} \psi_{1/2}$. These meromorphic 
zero modes $\psi_{1/2}$ are
well defined on the whole spectral cover apart from possible
singularities at the interaction points where the metric vanishes.
Their images under the boson-fermion
 correspondence yield $(2g_S-2+p)$  multi-valued ghost zero modes\footnote{ One can
 also find $g_S$ meromorphic zero modes obtained from the $g_S$ holomorphic differentials. 
The integration over the ghost fields is defined by discarding, in a similar way to the integration over the gauge fields,
these $g_S$ meromorphic zero modes.}.

It is easy to  construct  the zero modes $\psi_{1/2}$ in the case of an even spin structure.
One can find for each of the $2g_S-2+p$ interaction points a single
zero mode with a simple pole with residue one located at one of
the interaction points $I_i$, the Szego kernel $S_{I_i}(\lambda)$. This leads to  a $2g_S-2+p$ family of ghost
zero modes.  These zero modes $\psi_{\frac{1}{2}}$
are in one to one correspondence with the super-moduli of the spectral cover, i.e. 
 the holomorphic $3/2$ differentials
\begin{equation}
\psi_{\frac{3}{2}}=\omega_{\lambda}\psi_{\frac{1}{2}}.
\end{equation}
 The integration over the ghosts $(c,\bar c)$ is obtained by decomposing the zero modes
$c= \sum_a m_a \omega_{\lambda}^{-1/2}\psi_{\frac{1}{2}}^a$ where the $\psi_{\frac{1}{2}}^a$ form a basis
of the zero modes. The coordinates $m_a$ are the anticommuting super-moduli  coordinates. The measure of integration over
 the zero modes  is 
\begin{equation}
(\prod_a dm_ad\bar m_a
) {\vert \det (\psi_{\frac{1}{2}}^a,\psi_{\frac{1}{2}}^b)\vert}^{-1}
\end{equation}
where the integration is over the anticommuting variables $(m_a,\bar m_a)$.

The light cone Green-Schwarz action possesses a supersymmetry  
\begin{equation}
\delta_{\xi}x^I=\xi^{\alpha}\gamma^I_{\alpha \dot\alpha}\psi^{\dot \alpha},\\
\delta_{\xi}\psi_{\dot\alpha}=\xi^{\alpha}\gamma^I_{\alpha\dot\alpha}\bar 
 \partial x^I
 \end{equation}
where $\xi^{\alpha}$ is a  spinor.
In particular the variation of the action under this
supersymmetry reads
\begin{equation}
\delta_{\xi}S=-\frac{1}{\pi}\int d^2\lambda\partial \xi^{\alpha}J_{\alpha}
\end{equation}
where we have defined
\begin{equation}
J_{\alpha}=\gamma^I_{\alpha\dot\alpha}\bar\partial x^I \psi^{\dot\alpha}.
\end{equation}
The derivative
\begin{equation}
\chi^{\alpha}=\partial \xi^{\alpha}
\end{equation}
plays the role of a gravitino.
Similar expressions can be written with $J_{\dot\alpha}$.

The gravitino field is a classical field which encapsulates the
super- Riemann surface status. In particular the light-cone Green-Schwarz
action is written for $\chi=0$.  The deformation of the super-Riemann structure
induces a non-zero value for the gravitino field.
Consider a $(0,-1/2)$ anticommuting tensor $v$ which parameterizes
the deformation of the super-Riemann surface. Using a local basis
for the spinors $u^{\alpha}$, e.g. a local section of the bundle
$K^{1/2}_S\otimes S_+$ where $S_+$ is the vector bundle modeled
on the spin representation of $SO(8)$ is of the form
$\psi^{\alpha}= \psi\otimes u^{\alpha}$, we can define the supersymmetry parameter
\begin{equation}
\xi^{\alpha}=vu^{\alpha}
\end{equation}
The action is not invariant under the supersymmetry
transformations parameterized by $\xi^{\alpha}$.
The scattering amplitude is only invariant if the expansion of 
 the gravitino field is chosen to involve the supermoduli which
are explicitly integrated over
\begin{equation}
\chi^{\alpha}=(\sum_{i=1}^{2g_s-2+p} m_i \chi_i)\otimes
u^{\alpha}
\end{equation}
where the $\chi_i$'s define a slice of the
supermoduli space. They form a basis of the $(1,-1/2)$ tensors.
The coupling of the gravitino field to the supercurrent leads to a sum of terms in correspondence with each of the
interaction points
\begin{equation}
-\frac{m_i}{\pi}\int d^2\lambda \chi^i u^{\alpha}J_{\alpha}.  
\end{equation}
Integrating over the supermoduli is now easy, it amounts to
expanding the exponential of the action and picking the linear
terms in the supermoduli.
This leads to the following insertion at each of the interaction points
\begin{equation}
\int d^2\lambda  
\chi^i\bar \partial x^I\bar \Sigma^I
\end{equation}  
where we have define the world-sheet fermion with values in the
vector representation of $SO(8)$
\begin{equation}
\bar \Sigma^I=\gamma^I_{\alpha\dot\alpha}u^{\alpha}\psi^{\dot\alpha}.
\end{equation}
A convenient choice for the $\chi^i$'s is to be $\delta$-function
 supported\cite{verlin3}
leading to the insertion of
\begin{equation}
\partial x^I \Sigma^I \otimes \bar\partial x^I \bar\Sigma^I 
\label{pic}
\end{equation}
at the interaction point $I_i$. We have included
 the left and right sectors.
This is the operator insertion introduced in DVV.

Having identified the interaction generated by the integration over the supermoduli let us come back to
 the origin of the field $\Sigma^I$.
As defined it involves a local section of the $SO(8)$ spin bundle
$u^{\alpha}$.
Introducing the twist field
\begin{equation}
\sigma=\frac{1}{\sqrt{\omega_{\lambda}}}
\end{equation}
and  using the section $u^{\alpha}$ one can define the world-sheet spinor
\begin{equation}
\Sigma^{\dot\alpha}=\sigma u^{\dot \alpha}
\end{equation}
from which one can derive the operator product expansion
\begin{equation}
\psi^{\alpha}(\lambda) \Sigma^I(0)\sim \frac{\gamma^I_{\alpha\dot\alpha}}{\sqrt{\lambda}} \Sigma^{\dot \alpha}(0)
\end{equation}
in the neighbourhood of one of the interaction points.
Of course  this identifies the pair $(\Sigma^I,\Sigma^{\dot\alpha})$ with the spin fields associated to $\psi^{\alpha}$.

Finally let us comment on the ambiguities of the integration over the supermoduli.
The insertion of the picture changing operator is intimately linked to the choice of
 the basis $\chi_i$. Another choice for the $\chi_i$'s result in a different result
 after integration over the supermoduli. As in the case of superstring perturbation theory this
 ambiguity might only be resolved by a global definition of the integration over the supermoduli space.

\section{The Large $N$ Limit of the Scattering Amplitudes}

We have now identified the scattering amplitudes in matrix string theory with
a discrete version of the light cone string scattering amplitudes.
The discretization appears in the $x_-$ sector of the theory. Moreover we have reproduced the
DVV operator insertion at the string interaction points. In this section we consider the issue
of the integration measure over the instanton moduli and the link with the Weil-Petersson measure.

In the path integral we consider a background Abelian gauge field $a_S$ as defined in section 3 and expand
around this background configuration. One has to sum over all the possible
Wilson lines to take into account the different possible backgrounds. 
In the small coupling regime the contribution from the background gauge field decouples leaving 
a Gaussian integral defined
by excluding  from the path integration over the one-form $a_{\lambda}$  the zero
modes due to the $g_S$ holomorphic Abelian differentials on the
spectral cover. The result of the Gaussian integral is given by
the determinant of $\Delta_1$ acting on one-forms.
There is a correspondence $\phi_1 \to \phi_1/\omega_{\lambda}$ between the normalizable
eigenvectors $\phi_1$
of $\Delta_1$ and the normalizable eigenvectors $\phi_0$ of the Laplacian on scalars $\Delta_0$ for
non-vanishing eigenvalues.  This gives for the integral over $a_{\lambda}$
\begin{equation}
\frac{1}{\det' \Delta_0}
\end{equation}
The constant zero mode is not in  the Hilbert space  as it is not a normalizable mode.
Similarly the integral over the eight scalar coordinates leads to
a factor
\begin{equation}
\frac{1}{\det'^4 \Delta_0}
\end{equation}
Combining the two determinants we find a factor of
\begin{equation}
\frac{1}{\det'^5 \Delta_0} 
\end{equation}
The quadratic action of the ghost field leads to a Gaussian
integration over non-zero modes. 
The resulting integration yields the  determinant of the Laplacian operator acting on scalars
\begin{equation}
\det\ '(\Delta_{0}).
\end{equation}
Combined with the integration over the space time fermions we get
a factor of
\begin{equation}
 \det\ '^5(\Delta_{0}). 
\end{equation}
cancelling the bosonic determinants.

Let us now come to grips with the integration over the matter fields $x_d$ and $(\psi,\bar\psi)$.
Recall that the fermions $(\psi,\bar\psi)$ are world-sheet fermions on the spectral cover with
 values in the spin representations of $SO(8)$. We use the Mandelstam coordinates $w$.
Consider now a polynomial $ P(\partial x_d^I,\bar \partial x_d^I,\psi^{\alpha},\bar\psi^{\dot\alpha})$ and the following
vertex operator
\begin{equation}
V= P(\partial x_d^I,\bar \partial x_d^I,\psi^{\alpha},\bar\psi^{\dot\alpha})e^{ip^Ix^I_d}
\end{equation}
corresponding to the insertion of states $\vert phys>=V\vert 0>$ with momentum $p^I$ at  one of
 the marked points of the spectral cover.  
These vertex operators allow to define the string wave functions which are inserted at the end of the
 open cylinders of the  Mandelstam  diagram. 
The open cylinders are obtained  using  a cut-off limiting the length of the external cylinders. The ends of
 the  open cylinders are  located at times
$\tau_a$. 
Consider the wave function specified by the boundary  string configuration $(x^I_d=x_0,\psi=\psi_0,\bar\psi=\bar \psi_0)$ at
 the end of an external cylinder
\begin{equation}
\phi_a(x_0,\psi_0,\bar\psi_0,\tau_a)=
\int_{x_0,\psi_0,\bar \psi_0} {\cal D}x^I_d{\cal D}\psi_{\alpha}{\cal D}
\bar \psi_{\dot\alpha}V_a e^{-S(x^I_d,\psi_{\alpha},\bar\psi_{\dot\alpha})}
\end{equation}
 obtained by integrating over the fields defined on  the infinite cylinder $\tau\in
[-\infty,\tau_a]$  and the disc closing the cylinder at infinity
 where  the vertex operator $V_a$ is inserted.
The explicit dependence on the stopping times $\tau_a$ of the wave function $\phi_i(\tau_a)$ is
 removed by inserting the 
propagator\cite{gidd} 
\begin{equation}
\phi_a(x_0,\psi_0,\bar\psi_0)=e^{-2ip^-_a\tau_a}\phi_a(x_0,\psi_0,\bar\psi_0,\tau_a)
\end{equation}
where $(p^+_a,p_a^-,p^I_a)$ is the  momentum of the  
state inserted at the marked point $P_a$. The momentum $p^+_a$ appears explicitly in the construction of
 the Mandelstam  diagram, the momentum $p_a^-$ is explicit in the
wave function $\phi_a$ while $p_a^I$ completes the mass shell condition.
The scattering amplitude is obtained by integrating the wave functions $\phi_a$ over
 the space-time fields $(x^I_d,\psi_{\alpha},\bar\psi_{\dot\alpha})$ subject
to the boundary conditions $(x_0,\psi_0,\bar\psi_0)$ at the end of the external cylinders
\begin{equation}
\int  {\cal D}x^I_d{\cal D}\psi_{\alpha}{\cal D}\bar
 \psi_{\dot\alpha}(\prod_{i=1}^{2g_S-2+p}V_i) (\prod_{a=1}^p \phi_a) e^{-S(x^I_d,\psi_{\alpha},\bar\psi_{\dot\alpha})}= 
<\prod_{a=1}^p \phi_a(p_a)>
\end{equation}
where the DVV operators $V_i$ have been inserted at the interaction points
\footnote{ There is a vanishing factor coming from the Liouville action due to the non-chiral splitting of the Laplacian acting on fermions. It
 is given by $((\prod_i\rho(P_i))/(\prod_i \rho (I_i)))^{-1/3}$ where $\rho$ is the Weyl factor. This term has to be absorbed in
 the definition of the path integral}.

We still have to take into account the sum over the classical
instantons. This amounts to summing over the moduli space of the
instantons. Moreover we must divide by the number of twisted sectors due to the twisting by
Wilson lines. On the whole this leads to 
\begin{equation}
A=\frac{1}{N^{2g_S}}\sum_{k_1...k_{2g_S}}\int_{{\cal M}_R} d\mu
 {\vert \det (\psi_{\frac{1}{2}}^a,\psi_{\frac{1}{2}}^b)\vert}^{-1}
<\prod_{a=1}^p \phi_a(p_a)>
\end{equation}
where $d\mu$ is the measure of the instanton moduli space.

This measure specifies that one has to integrate over the coordinates of 
the interaction points of the Mandelstam diagram.  A change of
the parameters of the instantons amounts to moving the
coordinates of the interaction points. Let us describe the zero modes associated to the variation
of the interaction points, i.e. the tangent space to the moduli space\cite{t2}.
Locally around an interaction point the spectral cover $S_0$ is defined by
\begin{equation}
\tilde w_0= \frac{1}{2}\lambda_0^2
\end{equation}
 corresponding
to a one form  
 $\pi_0^* X_d=\lambda_0d\lambda_0$.
Let us perturb this spectral cover and move the interaction point. 
The equation of the perturbed $S_{\epsilon}$ reads
\begin{equation}
\tilde w_0=\frac{1}{2}\lambda(\lambda -4\epsilon).
\end{equation}
The interaction point is at $\lambda^*=2\epsilon$.
The one form associated to $S_{\epsilon}$ is 
 $\pi_{\epsilon}^*X_d=(\lambda-2\epsilon)d\lambda$.
It is easy to see that $(\lambda-\epsilon)d\lambda=\lambda_0 d\lambda_0$ implying that 
\begin{equation}
 \pi_{\epsilon}^*X_d=\pi_{0}^* X_d -\epsilon\phi
\end{equation}
 where $\phi=d\lambda$, i.e. the one form defining $S_0$ has been perturbed by $\phi$. 
Notice that the perturbation is holomorphic on $S_{\epsilon}$. This is not the case when
 pulled back to $S_0$ where it reads in the vicinity of the two singularities
\begin{equation}
\phi \sim  \pm(\frac{\pm i\epsilon/2}{\lambda_0 - (\pm i\epsilon)}))^{1/2}d\lambda_0.
\end{equation}
These infinitesimal perturbations of $S_O$ are determined by the one form $d\lambda_0/\lambda_0^{1/2}$ shifted away from the
interaction points by $\pm i\epsilon$. 
So the zero modes characterizing the tangent space to the moduli space are spanned by the
analytic one forms with a square root divergence at the interaction  points of the spectral cover.
Globally these one forms are  found to be
 $\sqrt{\omega_{\lambda}}\psi_{1/2}$
where $\psi_{1/2}$ is one of the $2g_S-2+p$  spinors with a pole at the the interaction points.
This identifies the tangent space to the moduli space and emphasizes the crucial role played
by the interaction points. In particular we retrieve the dimension of the moduli space $2g_S-3+p$
measuring the relative positions of the interaction points.

The measure 
in terms of the coordinates $ w_i$ of the interaction points is now
\begin{equation}
d\mu={\vert \det (\psi_{\frac{1}{2}}^a,\psi_{\frac{1}{2}}^b)\vert} (\prod_{i=1}^{2g_S-3+p}dw_id\bar w_i)
\end{equation}
This is the  measure on the flat cylinders forming the Mandelstam diagram.
Notice that the determinant arising from the tangent space to the moduli space cancels
the determinant coming from the supermoduli measure. 
Inserting this measure in $A$ we find that the  scattering amplitude is obtained by
integrating over all the possible topologies of the string diagram as specified by
the coordinates of the interaction points. 
 
The scattering amplitude $A$ on a spectral cover of genus $g_S$
carries a factor of $(1/g)^{2g_S-2+p}$ coming from the $2g_S-2+p$
fermionic zero modes. This springs from the rescaling $c\to gc$
used in the semi-classical expansion. Now as the string coupling
is inversely proportional to $g$ we find that the scattering
amplitude is weighted with a factor $g_s^{-\chi (S)}$ as it should in
the string scattering expansion\footnote{Here the Euler
characteristic $\chi(S)$ is defined for the open Riemann surface
with punctures obtained after removing a disc around the marked points.}.
 Notice that this amounts to adding  a factor of $g_s$
 to each of the interaction points.

So the scattering amplitude carries the appropriate factor of
$g_s$ to be identified with the string amplitude.
Together with the Hamiltonian picture of GS states propagating
on the Mandelstam diagram this shows that the scattering amplitude
of matrix string theory is  similar to a discrete version of the usual type IIA
superstring amplitude in the light cone gauge.

Let us now investigate the $N$ dependence of the scattering amplitude $A$. 
First of all the geometry of the spectral cover is independent of $N$. Indeed we have seen that 
the number of moduli is uniquely specified by the genus $g_S$ and the number
of marked points $p$. Fixing $g_S$ and $p$ in the large $N$ limit one can study the amplitude
 on a prescribed string diagram. Now the Mandelstam diagram depends on the momenta
and shifts due to the Wilson lines. In the large $N$ limit these momenta and shifts cover the entire real line,
the discretization step $1/N$ goes to zero. So we retrieve the light cone Mandelstam diagrams
of string theory. In particular all the external and internal momenta become continuous.
It is then easy to see that in the large $N$ limit the discrete sums in $A$ converge to the integral 
over the measure $\prod_{a=1}^{g_S}\alpha_a d\eta_a d\alpha_a $. On the
whole, combining with the measure $d\mu$,  we reconstruct the
  Weil-Petersson
measure over the moduli space of marked Riemann surfaces
\begin{equation}
d\hbox{ WP}=(\prod_{i=1}^{2g_S-3+p}\alpha_id\theta_i d\tau_i)(\prod_{a=1}^{g_S} \alpha_a
d\eta_a d\alpha_a )
\end{equation}
in terms of the twisting angles $\theta_i$ and $\eta_a$  (\ref{twist}).
The convergence towards the Weil-Petersson measure is only a weak convergence.

Having obtained the measure of the moduli space of curves in the large $N$
limit we find that the scattering amplitude $A$  
coincides with  the scattering amplitude of superstring theory in the light cone gauge.

\section{Conclusion}

We have given an explicit description of the moduli space of matrix string instantons. They are
 defined in terms of $(2g_S-3+p)$ continuous parameters in correspondence with the deformations of the spectral
 cover. The scattering amplitudes in matrix string theory are
equivalent to string scattering amplitudes on a Mandelstam
diagram where the interaction points are free to move while the
 momenta of the interacting strings are quantized due to the
presence of discrete Wilson lines in the description of 
 flat bundles. The interaction points are associated to the supermoduli describing the
 structure of Super-Riemann surface on the spectral cover. 
We have explicitly integrated over the supermoduli and retrieved
the DVV insertion of the picture changing operators at the interaction points.
Finally we have studied the large $N$ limit and shown that the measure
on the matrix string instanton moduli space  converges to the Weil-Petersson measure
on the moduli space of marked Riemann surfaces. 

\vskip 2 cm

\noindent Acknowledgements: I would like to thank D. Bernard for useful comments on vector bundles
and F. Nesti, P. Vanhove and T. Wynter for numerous discussions on matrix theory. This work was  partially supported by the EU Training and Mobility
of Researchers Program (Contract FMRX-CT96-0012).

\vfill\eject
\setlength{\baselineskip}{0.666666667\baselineskip}

\end{document}